\documentclass[sigconf]{acmart}

\AtBeginDocument{%
	\providecommand\BibTeX{{%
			\normalfont B\kern-0.5em{\scshape i\kern-0.25em b}\kern-0.8em\TeX}}}

\usepackage{algorithmic}
\usepackage{graphicx}
\usepackage{textcomp}
\usepackage{xcolor}
\def\BibTeX{{\rm B\kern-.05em{\sc i\kern-.025em b}\kern-.08em
		T\kern-.1667em\lower.7ex\hbox{E}\kern-.125emX}}
\usepackage{tikz}

\usepackage{caption}
\usepackage{subfig}

\usepackage{enumitem}
\usepackage{url}

\newcommand{\aname}{BarrierBypass }
\newcommand{\anametwo}{BarrierBypass}

\renewcommand\footnotetextcopyrightpermission[1]{}
\setcopyright{none}

\begin{document}


\title{BarrierBypass: Out-of-Sight Clean Voice Command Injection Attacks through Physical Barriers}


\author{Payton Walker}
\email{prw0007@tamu.edu}
\affiliation{%
	\institution{Texas A\&M University}
	\city{College Station}
	\state{Texas}
	\country{USA}
}

\author{Tianfang Zhang}
\email{tz203@scarletmail.rutgers.edu}
\affiliation{%
	\institution{Rutgers University}
	\city{New Brunswick}
	\state{New Jersey}
	\country{USA}
}

\author{Cong Shi}
\email{cs1421@scarletmail.rutgers.edu}
\affiliation{%
	\institution{Rutgers University}
	\city{New Brunswick}
	\state{New Jersey}
	\country{USA}
}

\author{Nitesh Saxena}
\email{nsaxena@tamu.edu}
\affiliation{%
	\institution{Texas A\&M University}
	\city{College Station}
	\state{Texas}
	\country{USA}
}

\author{Yingying Chen}
\email{yingche@scarletmail.rutgers.edu}
\affiliation{%
	\institution{Rutgers University}
	\city{New Brunswick}
	\state{New Jersey}
	\country{USA}
\vspace{15mm}
}

\begin{abstract}
	The growing adoption of voice-enabled devices (e.g., smart speakers), particularly in smart home environments, has introduced many security vulnerabilities that pose significant threats to users' privacy and safety. When multiple devices are connected to a voice assistant, an attacker can cause serious damage if they can gain control of these devices. 
	We ask where and how can an attacker issue \textit{clean} voice commands stealthily across a \textit{physical barrier}, and perform the first academic measurement study of this nature on the command injection attack. We present the \aname attack that can be launched against three different barrier-based scenarios termed \textit{across-door}, \textit{across-window}, and \textit{across-wall}. We conduct a broad set of experiments to observe the command injection attack success rates for multiple speaker samples (TTS and live human recorded) at different command audio volumes (65, 75, 85 dB), and smart speaker locations (0.1-4.0m from barrier). 
	
	Against Amazon Echo Dot 2, \aname is able to achieve 100\% wake word and command injection success for the across-wall and across-window attacks, and for the across-door attack (up to 2 meters). At 4 meters for the across-door attack, \aname can achieve 90\% and 80\% injection accuracy for the wake word and command, respectively. Against Google Home mini \aname is able to achieve 100\% wake word injection accuracy for all attack scenarios. For command injection \aname can achieve 100\% accuracy for all the three barrier settings (up to 2 meters). For the across-door attack at 4 meters, \aname can achieve 80\% command injection accuracy. 
	Further, our demonstration using drones
	yielded high command injection success, up to 100\%. 
	Overall, our results demonstrate the 
		potentially devastating nature of this vulnerability to control a user's device from outside of the device's physical space, and its limitations, \textit{without} the need for complex and error-prone command injection.
\end{abstract}


\maketitle
\pagestyle{plain}

\section{Introduction}
\label{sec:intro}

As voice assistant (VA) devices such as Amazon Echo and Google Home smart speakers are approaching ubiquity, we are forced to become more aware of the inherent security risks associated with these devices. VA devices typically act as a central hub of control for a multitude of connected smart devices such as smart locks, lights, cameras, thermostats, appliances, and garage doors. Each of these devices can be controlled in some way by issuing voice commands to the VA device. 
But these commands also introduce new types of risks. The ability to control such devices with vocal commands opens up a lot of attack possibilities that did not exist before. Among the different types of attacks that can be performed, the potential for home/office/hotel/dorm intrusion is one of the most severe and threatening.

Media coverage on this subject reveals the growing concern for the security vulnerabilities in a smart home environment~\cite{rambus,byteant,trendmicro,bobvilla,techspective}. While much of the concern is centered around the vulnerability to hacking that comes with connecting a multitude of devices, many professionals agree, for the purposes of home intrusion, there is a very low chance that an attacker would attempt to perform complex hacking as opposed to simply brute forcing there way in~\cite{wirecutter}. However, the ability to issue simple vocal commands to a voice assistant in order to control a lock or door is one vulnerability that requires no hacking and could potentially be favored by attackers who want to gain access to a space. 

Aside from commands being accidentally issued through television advertisements~\cite{broadcom,lifehacker,inc}, 
in an ISTR special report from Symantec, the author discusses the "mischievous man next door attack" which involves a neighbor issuing voice assistant commands either with ultrasonic frequencies, or by waiting until you leave and simply shouting a command through the door~\cite{symantec}. The report touches on the significant security risk that is introduced if you have smart locks or a garage door that can be controlled by your voice assistant because it would allow an attacker to gain entry into your home. While home invasion is a serious concern when a command injection attack is possible, it is important to note that there are many other scenarios that can cause harm if an attacker can control the smart devices of a home. For example, turning on the stove can cause a gas leak or become a fire hazard. 

Another form of command injection attack that has emerged in academia in recent years is hidden voice commands that obfuscate command audio so it is unrecognizable to humans,
but recognizable by VA devices. However, hidden voice command attacks have limited applicability and their accuracy is generally low. They are also very sensitive to noise because of how specially they are crafted to begin with. Also, even after the past several years of research on these attacks \cite{Abdullah2021FaultIOA}, the vendors have not really come up with defenses to such attacks. This is perhaps because the vendors are likely ignoring them as being rather impractical or uneventful. Indeed, the recent work by Abdullah et al.~\cite{Abdullah2021FaultIOA} revealed that many of the hidden voice command attacks presented in research are not truly feasible in real-world settings due to their low accuracy and lack of transferability to different systems. 
%
Another recent work on command injection by Sugawara et al.~\cite{Sugawara2020LightCL} introduces the LightCommands attack which uses laser-based injection of the audio signal. The main drawbacks to this attack are that it requires a line of sight, is very complicated to setup/launch, and can be error prone. 

In this paper, we aim to address most of the aforementioned problems with the existing voice command injection attacks from the literature or practice. We focus on an attack model, \anametwo, in which a loudspeaker issues \textit{clean vocal commands} --- through a physical barrier --- to a voice assistant or other voice controllable technology that is located inside a home, office, or hotel room. 
While this attack model eliminates many of the complications of hidden command injection, it does introduce its own limitations. For example, because this attack injects clean commands and requires louder volumes, the attack would likely only be launched in certain scenarios such as when the user is not present in the space (such as during work hours).
We consider three different barrier types which serve as the entry points for the attacker to inject such out-of-sight voice commands:
\vspace{-2mm}
\begin{enumerate}[wide]
	\item \textit{\textbf{Window Barrier}}: The attacker injects a command through a window to target a voice assistant in the room. 
	The attacker can launch this attack in-person or remotely-controlled via drone technology 
	which can target multiple homes in a neighborhood or even high rise buildings with condos or offices. We demonstrate the feasibility of both attack scenarios in this work.
	\item \textit{\textbf{Door Barrier}}: The attacker injects a command through a door that connects to the space with the victim voice assistant. This barrier is likely most susceptible due to the \textit{thin gap} beneath the door above the flooring which is sufficient for the sound waves to pass into the space easily.
    \item \textit{\textbf{Wall Barrier}}: The attacker is located in an adjacent space and injects a command through an interior wall. This barrier is applicable to housing setups such as dorms, hotels, or apartments where adjacent units share a wall.
\end{enumerate}

\textit{Is issuing voice assistant commands across a physical barrier possible? What types of barriers can be attacked? How can such an attack be achieved and what particular settings are required in order to bypass the barrier? What are the limitations of this attack?} These are the main research questions that we consider during this work and seek to answer. We perform extensive experimentation to evaluate the \aname attack in different parameter settings such as command audio loudness and location of the voice assistant device to determine when this type of command injection attack is possible in a real-world scenario. To our knowledge, a broad study on the voice assistant command injection attack, across physical barriers, has yet to be conducted in academia. This work demonstrates when the \aname attack is practical and it can be used to inform future research directions on the subject.

\noindent \textbf{{Main Contributions and Results:}} We summarize our key contributions and results below: 
 
\noindent \textbf{\textit{(1).\ Design of Clean Voice Barrier-based Attacks:}}
We designed three different barrier-based command injection attacks to represent common materials/objects that may act as a physical barrier between an attacker and the victim's voice assistant during a command injection attack. Specifically, we define the \aname attack in the \textit{across-door}, \textit{across-window}, and \textit{across-wall} scenarios and assess the effect of each barrier type on the attack's success. 
We present an attack that circumvents the sophistication and complexity of hidden voice command or laser-based command attacks, achieving the same goal with high accuracy in certain scenarios.
	
\noindent \textbf{\textit{(2).\ Measurement Study Evaluating the Effect of Multiple Parameters:}}
We present a measurement study and conduct an array of experiments to evaluate the effect of different barriers, under different attack settings, on command injection attack success. 
We test different speakers, loudness levels, voice assistant models (Amazon Echo Dot 2 and Google Home mini), device distances from the barriers, and observe the effect of different across-wall constructions (with and without insulation). 
\aname is able to achieve 100\% injection accuracy for both the wake word and command under certain conditions and selecting the highest performing speaker. 


\noindent \textbf{\textit{(3).\ Demonstration of Drone-based Attack:}}
We utilized two drone models equipped with Bluetooth speakers to demonstrate the potential for executing the \aname attack via drones. Our experimental simulations of the attack reveal high command injection success when using a drone that has a low operating loudness, or when the command audio is increased by 10 dB to compensate for a higher operating loudness. 

\noindent \textbf{\textit{(4).\ Informed Suggestions to Increase Attack Robustness and Defense Potential:}} 
Compiling the knowledge gained from our multiple experiments and attack demonstrations, we devise a set of suggestions that could be applied by an attacker in order to improve the potential for this attack under realistic conditions. Conversely, this information can be used to inform defensive mechanisms.
\section{Background}
\label{sec:background}

\subsection{Sound Passage Through Barriers}
As sound waves hit a physical barrier, they will lose energy and attenuate as they pass through the solid material. This occurs because the sound is either reflected off of the material (causing echo) or absorbed by it. Therefore, sounds on one side of a barrier played at a particular loudness (decibel) level, will be quieter when heard or recorded on the other side because the decibels are reduced. The transmission loss of sound across a barrier can be affected by many factors attributed to the barrier's material and construction. Thickness, density, and air space within the barrier are all factors that can either increase or decrease the level of sound transmission. For example, in double paned windows, thicker glass and greater air space in the middle are desired to optimize sound blockage~\cite{stc_window}. A barrier's ability to block sound is measured using different rating values such as Sound Transmission Class (STC) and Noise Reduction Coefficient (NRC).
We provide further detail on these values and what they represent in the following subsections.

\begin{figure*}[ht]
	\captionsetup{font=footnotesize,labelfont=bf,width=0.8\linewidth}
	\centering
	\includegraphics[scale=0.50]{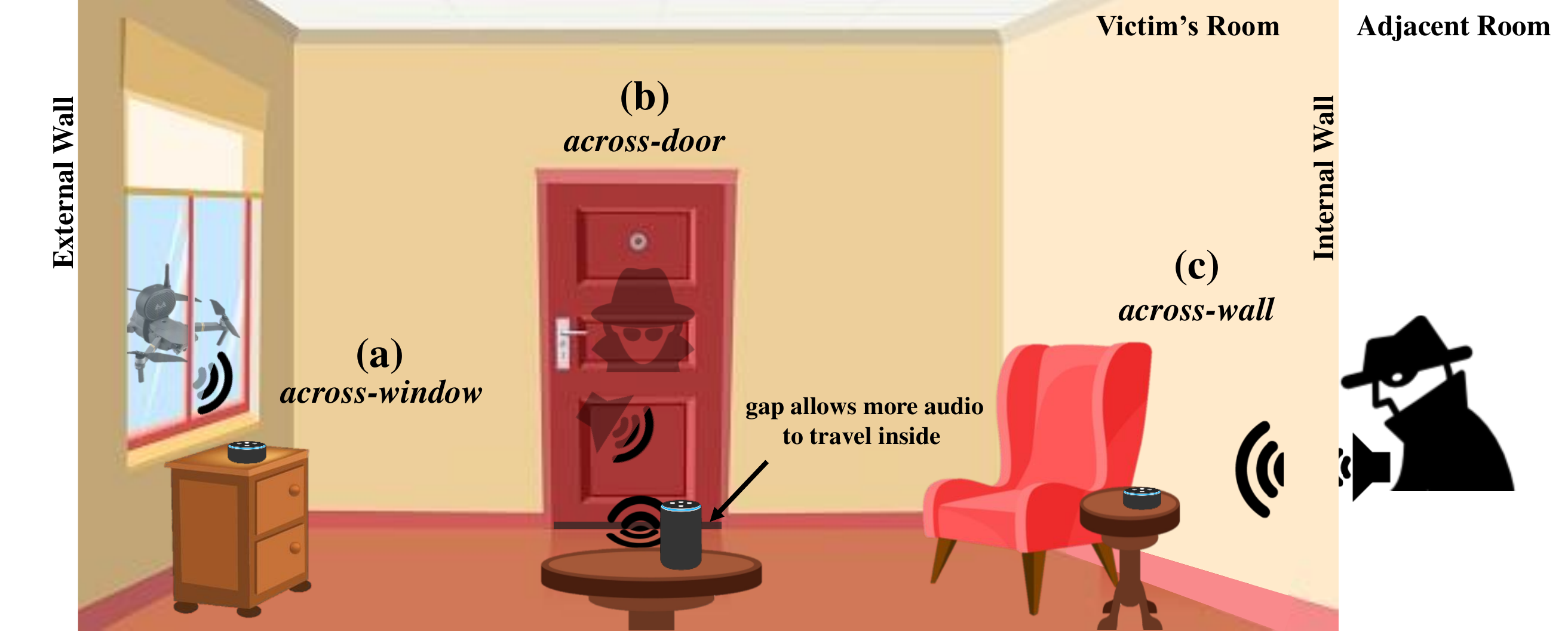}
	\caption{The \aname attack in the three barrier-based scenarios that we explore in our study including the (a) window barrier, (b) door barrier, and (c) wall barrier. The attacker is located on one side of the barrier, either in person such as an adjoining room or remotely using a drone, and attempts to inject an audible command to control the voice assistant located on the other side.}
	\label{fig:scenario_fig}
\end{figure*}

\subsection{Rating Values for Sound Propagation}

\noindent \textbf{Sound Transmission Class:}
Sound Transmission Class (STC) is an established rating system for how much sound is blocked by a particular assembly~\cite{stc}. It is an integer rating that roughly equates to the dB reduction in sound across a particular barrier. For example, a wall that reduces a 100 dB noise on one side, to a 60 dB noise on the other side would have an STC rating of 40. It is the most commonly used metric in the US for describing sound blockage potential and allows for direct comparison between different products (i.e., walls, doors, windows, etc.) and manufacturers. Specifically, the STC rating is calculated as the average noise blockage, in dB, for 18 different frequency values and has a logarithmic scale. This rating is based on the ASTM E413-16 standard~\cite{stc_standard}.

Since our work is mostly concerned with the amount of sound that is able to persist through a barrier and into the space on the other side, the STC rating is most relevant. The STC ratings for the different barrier setups that we consider include: STC of 20 for the door-barrier~\cite{stc_door}, STC of around 33 for the window-barrier~\cite{stc_window}, and STCs of 30 and 34 for the wall-barrier without insulation and with insulation, respectively. We will revisit these values later on when interpreting our experimental results. 

\noindent \textbf{Noise Reduction Coefficient:}
The Noise Reduction Coefficient (NRC) measures the amount of noise that a material absorbs~\cite{nrc}. Where the STC is a rating that describes how much noise can pass through a barrier, the NRC describes the amount of noise that is left within a space. Therefore, two materials with the same NRC does not imply that the same amount of noise is transmitted through the other side for each of them. NRC values are on a scale of 0 to 1, where 0 indicates the material will reflect back all of the sound that hits it, and a value of 1 indicates that all of the sound is absorbed by the material (e.g., none of it is reflected back). The NRC provides a single-value approximation of the noise absorption of a material by averaging the sound absorption coefficient values at four 1/3 octave frequencies (250, 500, 1000 and 2000 hertz) and is rounded to the nearest 0.05 increment. This rating is based on the ASTM C423-17 standard~\cite{nrc_standard}.


\section{Attack \& Threat Model}
\label{sec:attacks}

In this section we define three \aname attacks based on different types of barriers (Door, Window, Wall), depicted in Figure~\ref{fig:scenario_fig}, as well as describe our threat model.

\subsection{Barrier-Based Attacks}

\noindent \textbf{Across-Door Attack:}
The first barrier that we consider is a standard interior door. We define the \textit{across-door} attack to represent all situations where an attacker may attempt to inject a command to a victim's VA that is located across an interior door. If the door is locked, hindering the attacker from gaining direct access into the room, there is still the potential for the attacker to issue a command across the door barrier in order to achieve their goal (i.e., unlock the door's smart lock or control some other connected smart device). In this situation, the gap that exists between the bottom of the door and the floor can be considered a vulnerability that may be exploited by this attack. The presence of a small gap will significantly increase the audio propagation in the room and increase the potential for attack success.

\noindent \textbf{Across-Window Attack:}
The next barrier that we consider is a standard window. We define the \textit{across-window} attack to represent the more likely attack situation that an attacker is attempting to issue a command from outside the victim's home or office. Often the attacker will have no access to desired space, or even to an adjacent room, so issuing a command through a window may be their only option. Again, if the user can issue a command from this location, they may be able to gain access by issuing commands to other smart devices that are linked to the voice assistant (i.e., smart locks on the doors, smart garage door). The window used in our experiments was a builder's grade, double-pane window that was located on the balcony of a third floor apartment.

\noindent \textbf{Across-Wall Attack:}
The last barrier that we consider in this study is an interior wall. We define the \textit{across-wall} attack to represent the situations where an attacker may be in an adjoining room. This would be a common barrier for attackers in adjacent living arrangements such as apartment complexes, dorm rooms, or hotels. An attacker could easily set up the speaker equipment for their attack in their own space next door and not be disturbed. To allow for greater experimental control, we decided to simulate the across-wall scenario using a soundproof box and wall inserts that we constructed. We consider two typical constructions of interior walls that are still present today 1) without insulation and 2) with insulation. The details on the construction of the soundproof box and the wall inserts are provided in Sections \ref{subsec:soundproof_box} and \ref{subsec:wall_inserts}, respectively.

\subsection{Threat Model}
\label{subsec:threatmodel}


In our threat model, the attacker does not need prior knowledge of the target VA device or its settings. Through a process of initial testing with different wake words, an attacker can learn what device is in the victim space and how to activate it (e.g., the Amazon Echo only has four possible wake word settings so each could be tested). Also, depending on the placement of the target device, the attacker could look through a window of the target room (either in person or automated with a camera) and identify the device that is being used.
The attacker is equipped with a portable loudspeaker device that is pre-loaded with some voice commands that they would like to issue. The command audio can be recorded by the attacker themselves, generated using Text-to-Speech software, sourced from publicly accessible repositories of human speech samples, or recorded/synthesized samples of the victim's voice. Since modern voice assistants are not voice specific by default, the attacker does not necessarily need command audio that is in the victim's voice, making this attack easier to conduct. In fact, the attacker can run initial testing in their own space to identify a particular voice sample that performs the best for targeting a specific voice assistant device or passing through a specific barrier. 
\aname is designed as an untargeted attack that can be executed independent of the victim. The attacker can use any speech audio so there is no dependence on acquiring the user's speech. Therefore, the same attack setup can also be launched against many different victims successively in a short period of time. There is also a lot of freedom for the attack to target any available barrier separating them from the victim voice assistant (i.e., they can issue the command across all available windows or walls). In particular, the attack could launch \aname remotely using a drone device equipped with a loudspeaker. The drone can fly around to inject the command audio and could target all the windows in a home and even multiple homes (i.e., an entire street or neighborhood) and "leave" the scene very quickly if they suspect detection. They could also target apartments/condos in a high-rise building by flying the drone up to a window. Drones can be purchased cheaply and can come already equipped with a speaker~\cite{drone_speaker1} for \$150, or the speaker device can be purchased by itself for \$50~\cite{drone_speaker2} and attached to any drone.
While the \aname attack is fully functional as an untargeted attack, there is some potential for a more targeted approach against a specific victim. Using a replay or synthesis attack, an attacker can fool speaker recognition on a virtual assistant device and achieve even more severe attack capabilities.

While \aname is intended to be launched when the user is not home, there are some scenarios where it can be launched with the victim present in the space.
Because the command audio loses a lot of power and becomes quieter as it passes through a physical barrier, there is potential for the injected command to go unnoticed. In some cases the victim may be occupied doing some task or activity that may draw their attention away from their voice assistant (i.e., taking a shower, napping, watching TV in another room). During these times, the attack can still launch the attack successfully while avoiding detection.

Since the goal of the attack is to issue a command, we consider both parts of a voice assistant command audio, the wake word and the command itself. We recognize that wake word injection is foremost crucial for the attack because it activates the device to accept commands.
Additionally, injecting the wake word alone can open up new attack possibilities. When a voice assistant is woken up, a recording is made that is sent over the internet for processing and is typically stored in a command history log. Therefore, an attacker could inject the wake word with the intent of allowing the device to make an unauthorized recording of the audio in the space (i.e., user speech, audio from a television, music playing). The attacker may then compromise the online repository of VA recordings to learn private user information. \textit{While we evaluate the \aname attack on voice assistant devices, it is important to note that the attack is applicable to any voice controllable system.}

\section{Methodology}
\label{sec:methodology}

\subsection{Experimentation}
\label{subsec:exps}

\noindent \textbf{Parameters:}
To generalize the results from our experimental attack simulations, we consider multiple parameters and values. Aside from the three types of barriers and different setups for each, we also test VA command audio samples from Male and Female speakers that are generated using text-to-speech or recorded from live human speakers. We consider different loudness levels for the injected audio including 65 dB to represent normal conversational loudness, 75 dB to represent loud speech, and even 85 dB for very loud audio achievable using a loudspeaker device. 
We tested different distances of the VA device from the barrier including 0.1 and 0.5 meters for the across-window attack, and 0.1, 0.5, 1, 2, and 4 meters for the across-door attack. Lastly, we ran experiments using two different types of VA smart speakers.

\noindent \textbf{Experimental Setup:}
For each experiment we recreate a realistic attack setup with the portable loudspeaker placed on one side of the barrier (attacker side), and the target smart speaker on the other side of the barrier (victim side) at certain distances. We ensure that the loudspeaker and smart speaker devices are aligned directly across from each other with the loudspeaker facing the barrier. We use the digital sound level meter on the attacker side to set the SPL of the command audio from the loudspeaker to the appropriate loudness. As a representative example of a command an attacker may attempt to issue, we selected the single-word, "Disarm" command. We consider the scenario where an attacker may be attempting to enter a victim's home and needs to disable the security system that is linked to their smart home environment (i.e., smart speakers). However, we believe our results are representative of other types of single-word commands. For each experimental parameter setting, we attempted the attack 10 times and recorded the number of successful injections of the wake word and the command portions. With 12 speaker samples, 3 SPL levels, 10 barrier/distance combinations, and 2 smart speaker devices, conducted a total of 7,200 attack simulations as part of our evaluation.

\noindent \textbf{Command Audio Samples:}
We created a set of command audio samples consisting of both Text-to-Speech (TTS) samples and recordings of Live Human (LH) speakers saying the single-word command, ``Hey Google/Alexa, Disarm''. This command represents an attacker's attempt to turn off a user's home security system so that the attacker may gain access. We do not make any claims that our results are representative of other single-word commands, but we do believe that more complex commands would make the attack more difficult. Specifically, we use samples from three Male speakers (M1-M3) and three Female speakers (F1-F3), for both sample types, for a total of 12 different speaker samples. The TTS samples were generated using a free online text-to-speech generator~\cite{tts_site}, and the LH speech samples were recorded directly from volunteers.
Prior to our experimentation, we confirmed that all of the command audio samples that we collected achieved 100\% recognition success in the non-malicious setting (when there is no ambient noise or physical barriers present).

\noindent \textbf{Equipment:}
In our experiments we use a cheap and low-end Sony SRS-XB2 portable loudspeaker to play the command audio. Notably, more powerful speakers can improve attack success. For the victim voice assistant, we use both the Amazon Echo Dot 2 and Google Home mini smart speakers. In order to ensure the command audio was played at the correct sound pressure level (SPL) we use a Rolls SLM305 digital sound level meter. Additionally, we built our own soundproof box and wall inserts for the across-wall scenario.

\subsection{Soundproof Box}
\label{subsec:soundproof_box}

For the across-wall attack, we construct a soundproof box in order to self contain the experiments in a highly controlled space that allows us to test different wall constructions. This approach allows to select specific building materials with sound blockage ratings that we know beforehand and to ensure that the command audio is only able to reach the VA device by passing through the wall. We found this approach easier than attempting to learn what materials were used in the walls of a real environment. To build the soundproof box (pictured in Appendix Figure \ref{subfig:box}) we followed the instructions outlined in \cite{soundproof_box}. We lined a cardboard box with foam board using 3M Super77 Spray Adhesive. Next, we added a layer of 1/4'' thick Dynamat Dynaliner (Self-Adhesive Sound Deadener). Lastly we added a layer of 3'' Acoustic Foam Egg Crate Panels using Auralex Foamtak adhesive spray. The different layers of the soundproof box are shown in Appendix Figure \ref{subfig:layers}.

\subsection{Wall Inserts}
\label{subsec:wall_inserts}

To experiment with different across-wall barriers, we constructed two wall inserts to fit inside the soundproof box, pictured in Appendix Figure~\ref{fig:inserts}. These inserts are constructed to the exact measurements that allow the insert to fit inside the soundproof box with a tight seal around all edges. 
Appendix Figure \ref{subfig:exp_aerial} shows the setup for the across-wall attack experiments.
The inserts were built 
with and without insulation~\cite{soundproofing_walls_ceiling}. Both inserts have a 2"x4" wood frame and are encased in 5/8" drywall panels that are cut to the exact dimensions of the frame. One of the inserts contains R13 Fiberglass insulation inside the stud frame, while the other insert was left empty. The stud frames were connected using 1 1/2" wood screws, and the drywall was attached with drywall glue and screws.
\section{Attack Results}
\label{sec:results}

\begin{table*}[ht]
	\captionsetup{font=footnotesize,labelfont=bf}
	\centering
	\caption{Command injection success rates, for attacking the Amazon Echo Dot 2, for each Barrier scenario. *Table is condensed to include only rows that showed some injection success.}
	\vspace{-2mm}
	\includegraphics[scale=0.58]{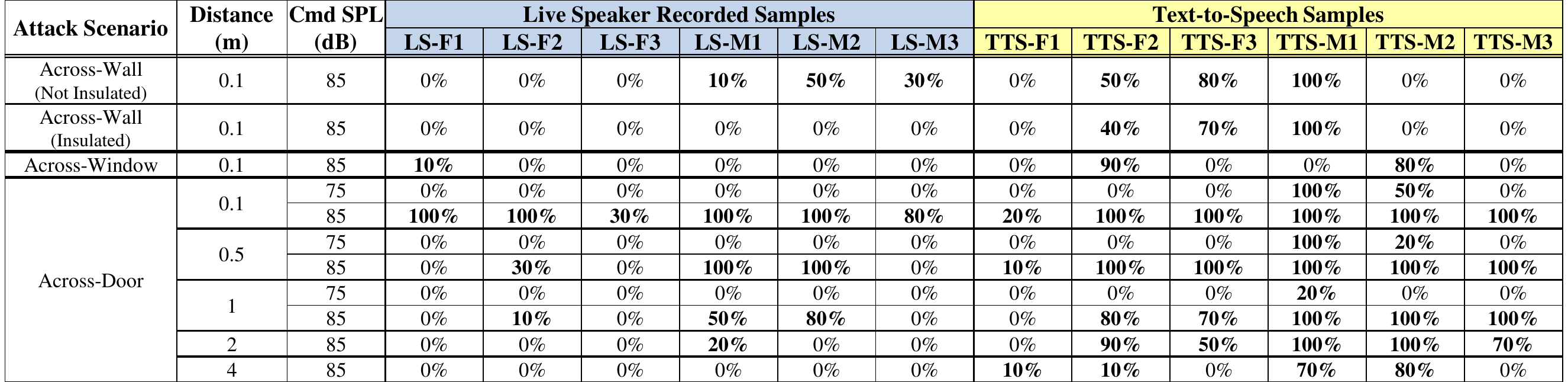}
	\label{tab:alexa-attack-results_cmd}
\end{table*}

\begin{table*}[ht]
	\captionsetup{font=footnotesize,labelfont=bf}
	\centering
	\caption{Command injection success rates, for attacking the Google Home mini, for each Barrier scenario. *Table is condensed to include only rows that showed some injection success.}
	\vspace{-2mm}
	\includegraphics[scale=0.58]{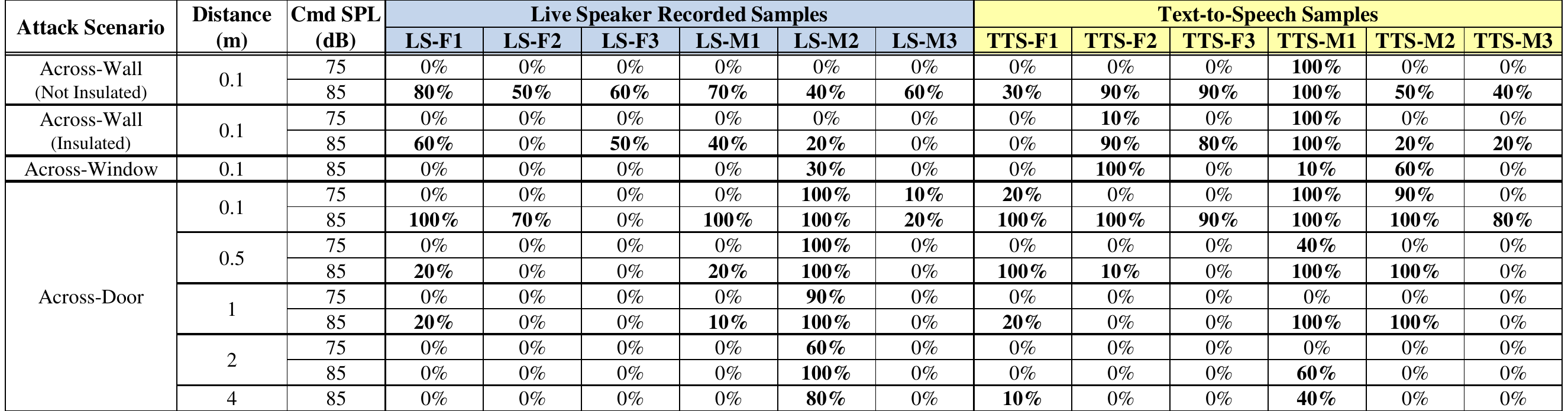}
	\label{tab:google-attack-results_cmd}
\end{table*}

In this section we report the \aname attack results from our experiments. 
We recorded and present both wake word and command injection success for all audio samples. 
Appendix Tables \ref{tab:alexa-attack-results_ww} \& \ref{tab:google-attack-results_ww} show the \textit{wake word} injection rates for the Amazon Echo Dot and Google Home mini smart speakers, respectively. And Tables \ref{tab:alexa-attack-results_cmd} \& \ref{tab:google-attack-results_cmd} show the \textit{command} injection rates. The values represent the percentage of successful injection out of 10 attempts. We present results for the standard implementation of \aname using non-specific voice audio for command injection, and discuss our investigation of the targeted implementation for fooling speaker recognition. To save space, we condensed the tables to include only the rows that showed instances of injection success. Therefore, any command SPLs or distances tested that were not included in these tables had no injection success for any of the speaker samples. 

\subsection{Standard \anametwo:}

\noindent \textbf{Across-Wall Attack:}
(Amazon Echo Dot 2) From our experiments for the across-wall attack, we observe that both wake word and command injection success was only possible when the audio was played at the loudest SPL level, 85 dB, when attacking the Amazon Echo Dot 2 device.  
If we compare the average injection success rates for both types of speakers for the across-wall attack with no insulation, we get 22\% success for the live speaker samples, and 50\% success for the TTS samples. And if we look at the across-wall attack with insulation we find that the wake word injection success completely diminishes to 0\% for the live speakers, and slightly decreases to 47\% for the TTS speakers. Comparing the injection success averages from the command injection results we see a similar trend. With no insulation, the live speaker samples have average injection success of 15\% and the TTS speaker have 38\%. And when insulation is added we again find the live speaker sample injection success drops to 0\% and the TTS speaker samples slightly decreases to 35\%. However, part of our threat model is that the attacker can perform preliminary testing and select the best performing command sample to launch their attack. \textbf{\textit{Choosing sample TTS-M1, the attack achieves 100\% success rates for wake word and command injection, at 85 dB for both types of walls, when targeting the Echo Dot .}}

(Google Home mini) For the Google Home mini we observed wake word and command injection success at 75 dB and 85 dB, and much greater success rates overall compared to the Amazon Echo Dot 2. Again, comparing the average wake word injection success rates for both types of speakers we find at 75 dB the average live speaker sample success is 95\%, outperforming the average TTS sample success of 68\%. At 85 dB, both live speaker and TTS samples achieve 100\% wake word injection success. When insulation is added the average success rates slightly decrease. At 75 dB, the live speaker and TTS sample success rates decrease to 85\% and 57\%, respectively. And at 85 dB the success rates decrease from 100\% for both speaker types with live speaker samples achieving 93\% and TTS samples achieving 72\%. Like the Amazon Echo Dot 2 results, we see a large decrease in command injection success compared to the wake word injection. At 75 dB, the live speaker samples had 0\% command injection success, and the TTS samples had 17\% command injection success. When the audio was played at 85 dB these average success rates increase to 60\% and 67\% for the live speaker and TTS samples, respectively. When the insulation was added, we see very similar success rates at the 75 dB level of 0\% and 18\% for live speakers and TTS samples, respectively. However, at 85 dB, we see a decrease in injection success (compared to no insulation) with live speaker samples dropping to 28\% and TTS samples dropping to 52\%. \textbf{\textit{Choosing sample TTS-F3 or TTS-M1 achieves 100\% success rates for injecting the wake word at both SPL levels, and TTS-M1 achieves 100\% success for injecting the command at both SPL levels, when targeting the Google Home.}}


\noindent \textbf{Across-Window Attack:}
In the across-window attack we observed injection success at 0.1 meters. Increasing the distance to 0.5 and 1 meter completely diminished injection success for all speaker samples and audio SPL levels. For both smart speakers we observed injection success at 75 dB and 85 dB for the wake word, and at 85 dB for the command.

(Amazon Echo Dot 2) 
At 75 dB we observe no wake word injection success for the live speaker samples, and only two instances of injection success (3\% average) for the TTS samples. When the audio was increased to 85 dB the average success rates increased to 15\% for the live speaker samples and 48\% for the TTS samples. No wake command injection success was observed at the 75 dB level, but at 85 dB we observed one instance of successful injection (2\% average) for the live speaker samples. The TTS samples showed greater success with an average of 28\% command injection success. \textbf{\textit{Selecting sample TTS-F2 or TTS-M1 allows the attack to achieve 100\% success rates for injecting the wake word at the 85 dB SPL level, and keeping TTS-F2 achieves 90\% success for injecting the command at the 85 dB SPL level, targeting the Amazon Echo Dot.}}

(Google Home mini)
In the results for the Google Home mini we observed nearly identical wake word injection success rates for the live speaker and TTS samples at both the 75 dB and 85 dB SPL levels. At 75 dB the live speaker samples had no wake word injection success and the TTS samples had only one instance of success (2\% average). When the SPL level was increased to 85 dB, both the live speaker samples and TTS samples showed an average of 78\% wake word injection success. Looking at the success rates for command injection, we find that the TTS samples were more successful. The live speakers samples had an average command injection success rate of 5\% while the TTS samples achieved 28\%. \textbf{\textit{Selecting any of the samples LS-F1, LS-F3, LS-M1, LS-M2, TTS-F2 or TTS-F3 achieves 100\% success rates for injecting the wake word at the 85 dB SPL level, and sticking with the TTS-F2 sample achieves 100\% success for injecting the command at the 85 dB SPL level, targeting the Google Home.}} 

\begin{table}[ht]
	\captionsetup{font=footnotesize,labelfont=bf}
	\centering
	\caption{Command injection success rates observed during the drone experiments.}
	\includegraphics[scale=0.35]{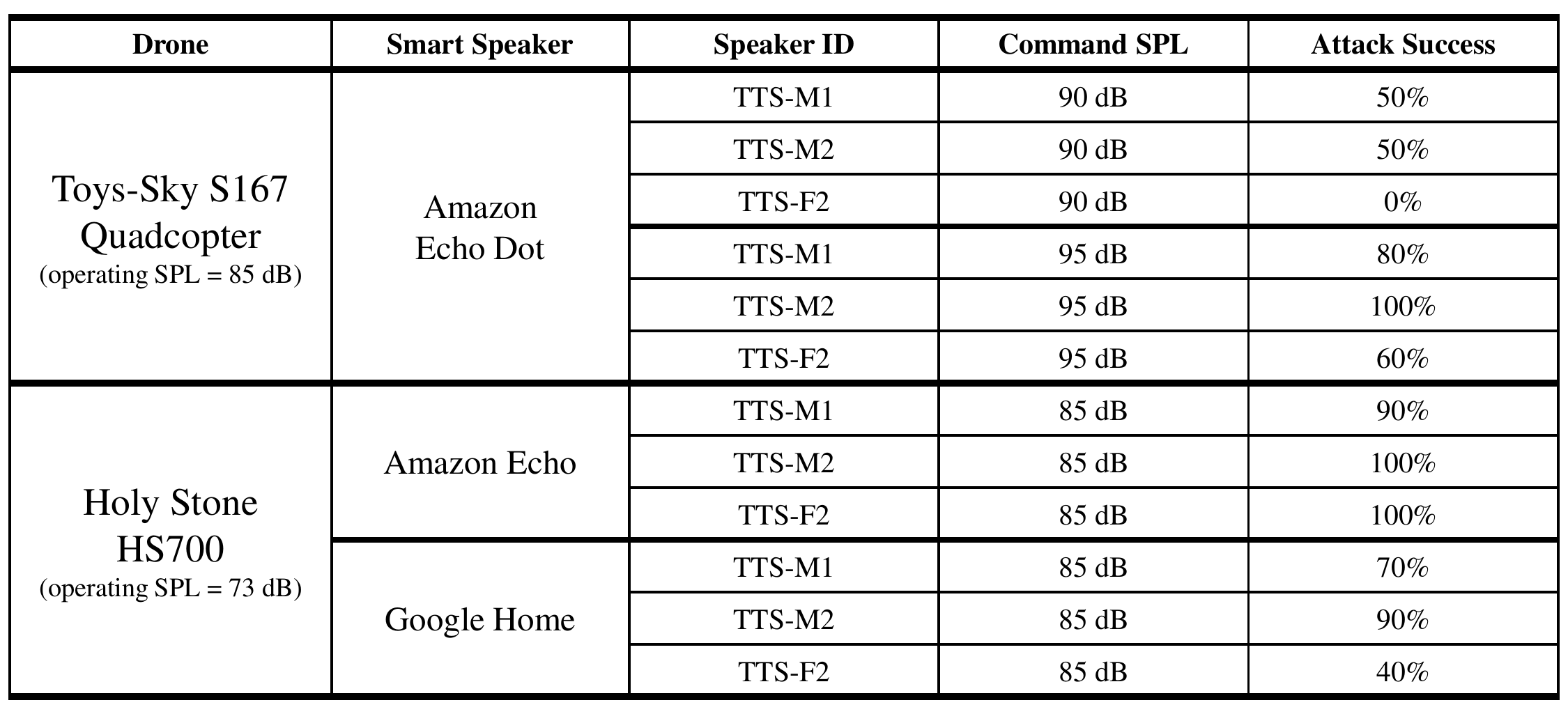}
	\label{tab:drone_results}
\end{table}

(Drone Attack)
Table~\ref{tab:drone_results} depicts the results for each drone-based scenario tested. Firstly, we found that using a drone with an operating loudness of 85 dB (S167) required command audio be played at 90+ dB. Specifically, at 90 dB we observed command injection success only up to 50\%. However, increasing the command audio to 95 dB allowed us to observe attack success up to 100\%. Since the operating loudness of the S167 was equal to the volume of audio used in our original experiments, the command audio in the presence of the drone had to be increase by at least 5 dB to overcome the added noise and achieve an SNR closer to 1.0 for successful command injection.

For our remaining experiments using the HS700 drone with a much lower operating loudness level, we observed high rates of command injection success, similar to what was observed in prior experiments when no drone was used. Because the operating loudness of the HS700 is only 73 dB, the 85 dB command audio level was not hindered by the added noise because it maintained a similarly high SNR. When targeting the Echo Dot, we observed attack success up to 100\%. And when targeting the Google Home we observed attack success up to 90\%.

\noindent \textbf{Across-Door Attack:}
In the across-door attack we observed wake word and command injection success rates at both the 75 dB and 85 dB SPL levels for most of the distances tested. Compared to other barriers, the results confirm that the door is easiest to compromise.

(Amazon Echo Dot 2)
For the live speaker samples, we observed wake word injection success at the 75 dB for the 0.1-meter distance only, achieving an average of 60\% injection success. At all other distances there was no wake word injection success at 75 dB. When the audio was raised to 85 dB, we observed a greater range of success across the different distances tested. On average, the live speaker samples achieved 97\%, 80\%, 62\%, and 12\% wake word injection success for the 0.1, 0.5, 1, and 2-meter distances, respectively. In comparison, the TTS samples showed greater success for both SPL levels and all distances. At 75 dB, the TTS samples achieved average wake word injection success rates of 55\%, 37\%, 30\%, and 17\% for the 0.1, 0.5, 1, and 2-meter distances, respectively. And at 85 dB, we observe wake word injection success rates of 100\% for 0.1 and 0.5 meters, and 88\%, 80\%, and 50\% for 1, 2, and 4-meter distances.

Looking at the results for command injection we again find decreased success rates compared to the wake word. For the live speaker samples we did not observe any command injection success at the 75 dB SPL level. At 85 dB, we observe average success rates of 85\%, 38\%, 23\% and 3\% for the 0.1, 0.5, 1, and 2-meter distances, respectively. Like the wake word results, we found that the TTS samples performed better for command injection. At 75 dB we observed average command injection success rates of 25\%, 20\%, and 3\% for the 0.1, 0.5, and 1-meter distances, respectively. And when the audio was raised to 85 dB, we observed average success rates of 87\%, 85\%, 75\%, 68\%, and 28\% for the 0.1, 0.5, 1, 2, and 4-meter distances, respectively. While multiple speaker samples showed very high success rates at certain SPL levels and distances, there are two samples that outperformed the rest. \textbf{\textit{By choosing TTS-M1 the attack can achieve 100\% success at both SPL levels up to 2 meter distances, and achieves 90\% success at the 4-meter distance. And choosing TTS-M1 or TTS-M2 for command injection allows the attack to achieve up to 100\% success rates at the 75 dB SPL level of distances up to 0.5 meters. When the SPL level is increased to 85 dB, the attack can achieve 100\% success for distances up to 2 meters, and 80\% success at 4 meters when launching the attack against the Amazon Echo Dot.}}

(Google Home mini)
We observed greater wake word injection success with the Google Home mini. For both 75 dB and 85 dB levels we see instances of wake word injection success at all distances that were tested. At 75 dB, the live speaker samples achieved average injection success rates of 85\%, 52\%, 53\%, 13\%, and 8\% for the 0.1, 0.5, 1, 2, and 4-meter distances, respectively. The TTS samples showed similar success rates of 67\%, 53\%, 30\%, 33\%, and 3\% for the 0.1, 0.5, 1, 2, and 4-meter distances, respectively. When the audio was increased to 85 dB, the average success rates increased. The live speaker samples achieved 100\% success for the 0.1, 0.5, and 1-meter distances, and achieved 78\% and 53\% for the 2 and 4-meter distances, respectively. The TTS samples achieved 100\% success for the 0.1 and 0.5-meter distances, and 98\%, 83\%, and 42\% success at 1, 2, and 4-meter distances.
 
For command injection, we observed a decrease in the average success rates for both speaker types. However, instances of success were still observed for both the 75 dB and 85 dB SPL levels for distances up to 2 meters. At 75 dB, the average command injection success rates for the live speaker samples were 18\%, 17\%, 15\%, and 10\% for the 0.1, 0.5, 1, and 2-meter distances. The TTS samples had less success at the larger distances and only achieved accuracies of 35\% and 7\% for the 0.1 and 0.5-meter distances. When the audio was played at 85 dB, both speaker types showed command injection success at all distances. The live speaker samples achieved average success rates of 65\%, 23\%, 22\%, 17\%, and 13\% for the 0.1, 0.5, 1, 2, and 4-meter distances, respectively. The TTS samples outperformed the live speaker samples at the shorter distances achieving success rates of 95\%, 52\%, 37\%, 10\%, and 8\% for the 0.1, 0.5, 1, 2, and 4-meter distances, respectively. \textbf{\textit{Choosing LS-M1, LS-M2, TTS-F1, or TTS-F3 will allow the attack to achieve 100\% success for wake word injection at both SPL levels up to 2-meter distances. At the 4-meter distance the attack can achieve 50\% and 100\% accuracy at the 75 and 85 dB levels, respectively. And isolating LS-M2 for command injection allows the attack to achieve up to 100\% success for both SPL levels up to 0.5 meters. At the 75 dB SPL level the attack can achieve 90\% and 60\% success for the 1 and 2-meter distances, respectively. And when the SPL level is raised to 85 dB the attack can achieve 100\% success at distances up to 2 meters and 80\% success at 4 meters when attacking the Google Home.}}

 \subsection{Targeted \anametwo}
 
 \noindent \textbf{Replay Attack:}
To investigate the potential for \aname to launch a replay attack across physical barriers, we performed a set of experiments using three speaker samples in the across-door attack setup. Specifically, we trained a voice profile for the LS-M1 speaker on both the Amazon Echo Dot 2 and Google Home mini. We recorded samples of the command "Alexa/Hey Google, what's my name?" from the live speakers LS-M1 and LS-M2, as well as a generated samples of the command using the text-to-speech speaker TTS-M1. We selected these speakers because they all achieved 100\% wake word and command recognition in the across-door attack. Playing each command audio at 85 dB we recorded the number of times out of 10 attempts that the voice assistant identified the trained speaker's voice.
We found that the Amazon Echo Dot 2 was 100\% accurate at identifying the trained speaker and denying the other speakers. For the Google Home mini we observed the device was 80\% accurate at identifying the trained speaker, and 100\% accurate at not identifying the untrained speakers. These experiments demonstrated that 1) speaker recognition can identify a valid user without a barrier present, 2) it will still accept a command from a random speaker (e.g., from attacker) across a barrier, and 3) it can identify a replayed voice of the valid user across a barrier. 

\noindent \textbf{Synthesis Attack:}
Synthesis attacks generate fake speech using a model trained on an original voice such that the synthesized voice matches the original. We performed another side investigation to observe the potential for successful synthesis attacks through physical barriers. We used the voice synthesis model SV2TTS~\cite{jemine2021RealTVC} from \cite{jia2018TransferLF} to generate the "Alexa/Hey Google, what's my name?" command in a live speakers voice. That same live speaker trained voice profiles on Amazon Echo Dot 1 and Google Home smart speakers. In the across-door attack setup, we played the synthesized command at 85 dB and recorded the number of times out of 10 attempts that the voice assistants identified the synthesized audio as coming from the legitimate user. We found that the synthesized commands were 100\% successful at fooling the speaker recognition function on both of the smart speakers. This further broadens the threat level and devastating potential of the \aname attack because it demonstrates that fake commands synthesized in a user's voice are sufficient enough to fool speaker recognition, even across physical barriers.

\section{Signal Analysis}
\label{sec:analysis}

\begin{figure*}[ht]
	\captionsetup{font=footnotesize,labelfont=bf,width=1\linewidth}
	\setlength{\belowcaptionskip}{-5pt}
	\centering
	\subfloat[0-8 kHz]{
		\includegraphics[scale=0.3]{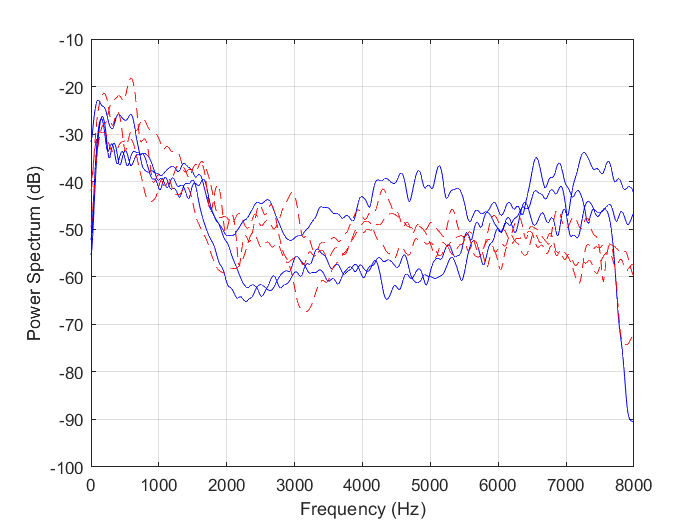}
		\label{fig:pspec_plots-a}
	}
	\subfloat[6-7 kHz]{
		\includegraphics[scale=0.3]{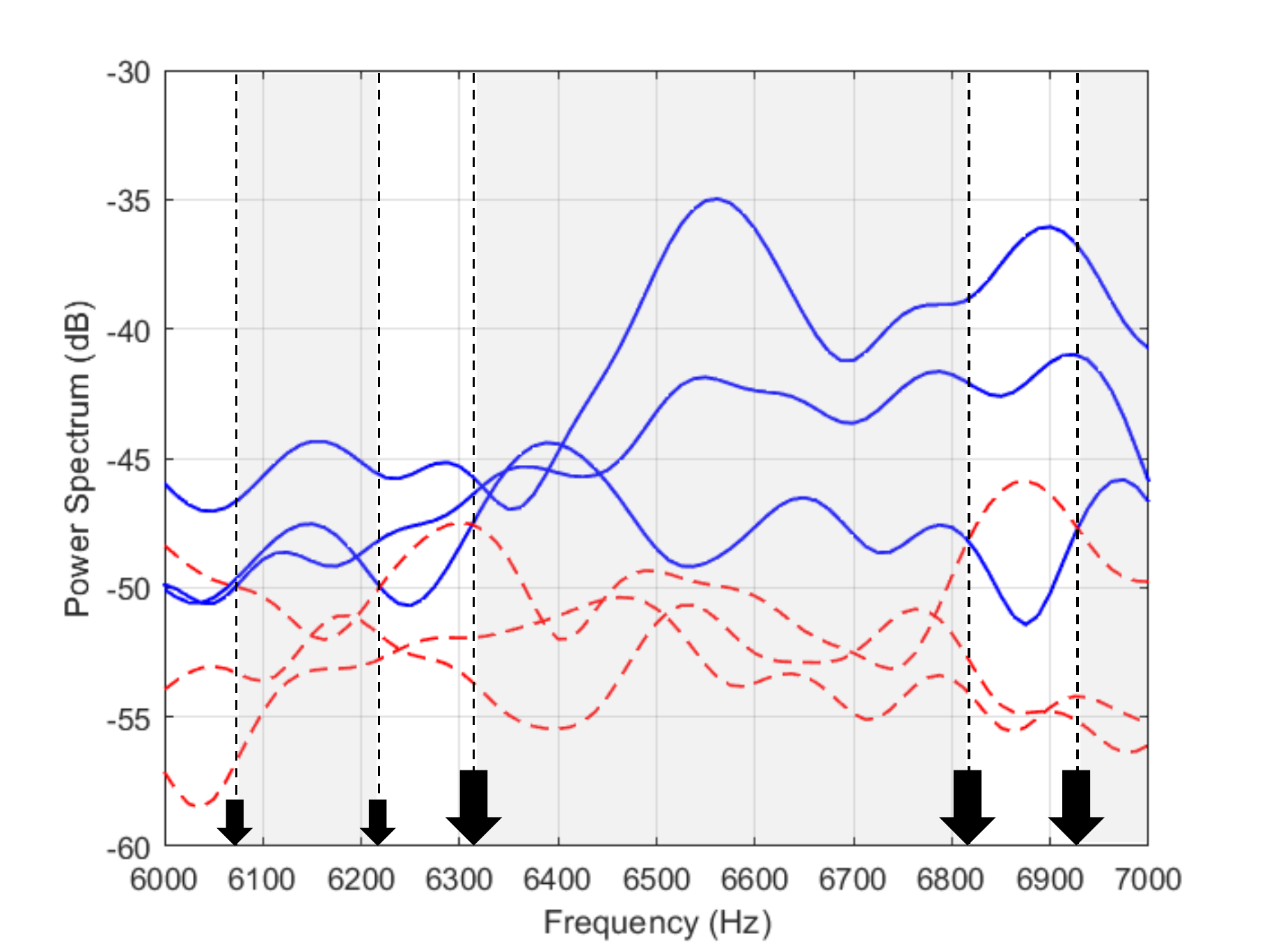}
		\label{fig:pspec_plots-b}
	}
	\subfloat[7-8 kHz]{
		\includegraphics[scale=0.3]{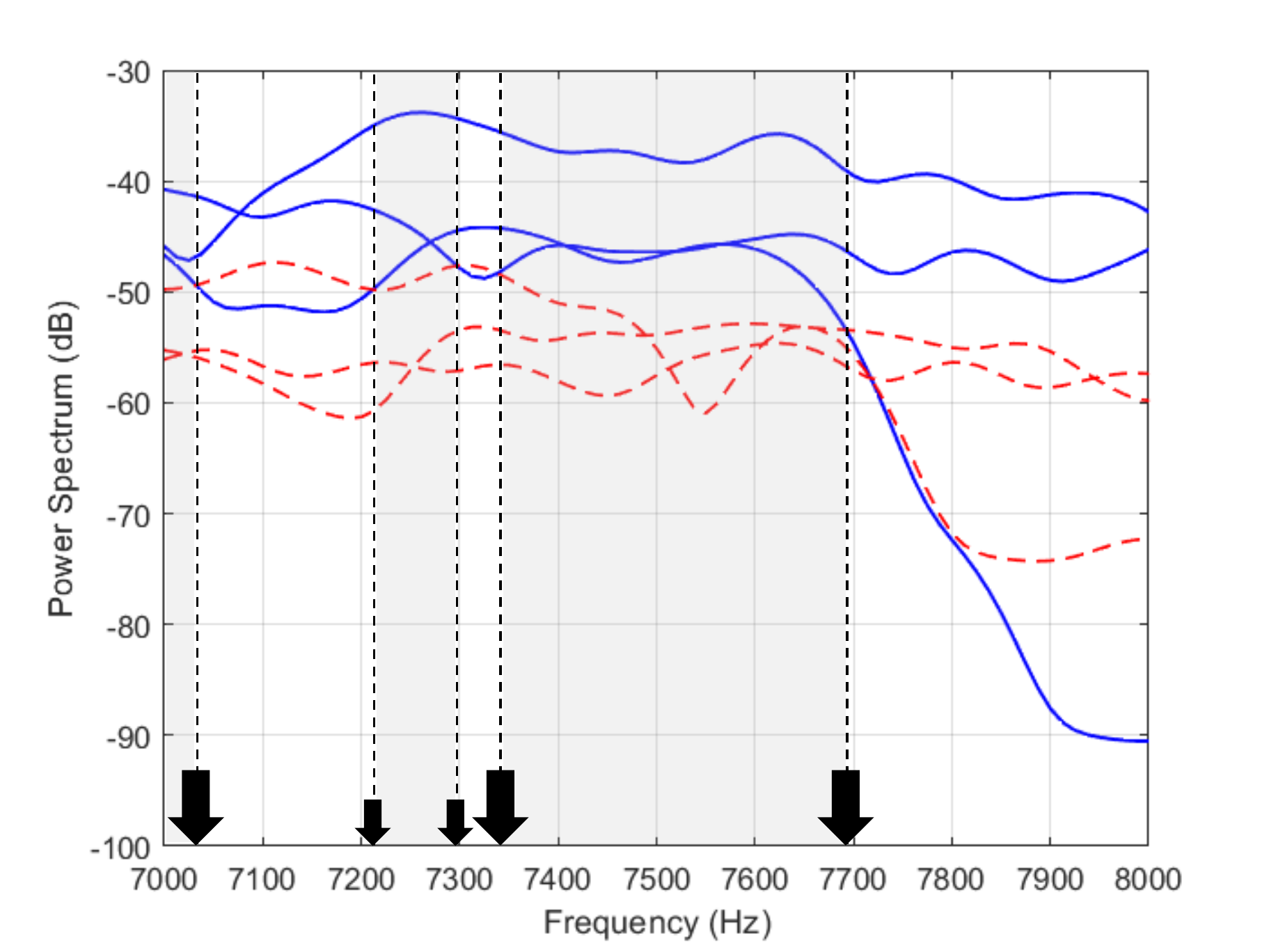}
		\label{fig:pspec_plots-c}
	}
	\caption{Power spectrum graphs of the wake word from each command audio sample that showed injection success (blue) and failure (red) in the across-wall scenario (without insulation).}
	\label{fig:pspec_plots}
\end{figure*}

In order to improve our understanding of why certain speech samples were successfully injected across the barriers
we investigated what frequencies were most affected by the barriers and whether we could identify certain frequency characteristics in our command audio samples that may explain the different levels of success. 

\noindent \textbf{Power Spectrum}
We generated power spectrum graphs that overlay the spectrums for each of the command audio samples, isolating the wake word specifically, in order to compare frequency distributions and identify specific characteristics. We chose to investigate the wake word portion of the commands because 1) command injection cannot occur unless the wake word is successfully issued, and 2) the injection attack experiment results showed greater success/failure distinction, for the wake word, between different speakers. 
In Figure \ref{fig:pspec_plots} we show the power spectrums of the wake word portion of the command audio samples for each individual TTS speaker. In the graphs, the solid blue lines indicate power spectrums of speaker samples that were successful at injection, while the red dashed lines indicate power spectrums of speaker samples that were not successful. From these graphs, we identify certain frequency characteristics that are consistent among the successful samples. Figure \ref{fig:pspec_plots-a} shows the full power spectrum of frequencies from 0 to 8 kHz. Looking at this graph we find there are certain frequencies in the upper range that have consistencies between the successful and failing samples. Figures \ref{fig:pspec_plots-b} \& \ref{fig:pspec_plots-c} show power spectrums that zoom into the frequency ranges of 6 to 7 kHz and 7 to 8 kHz, respectively. In these graphs we can identify five different frequency ranges (6.08-6.22 kHz, 6.32-6.82 kHz, 6.93-7.04 kHz, 7.21-7.30 kHz, and 7.34-7.69 kHz) where we find that all samples that showed successful injection have stronger frequencies in these ranges than the samples that were not successful. 

While more sophisticated exploration is needed to make final conclusions, we have a few hypotheses about why certain audio samples performed better than others. First, it is possible that audio samples that showed greater success utilized more bass in the part of the wake word that are required for recognition. Therefore, as the audio passes through the physical barriers and those components of the audio are strengthened, the audio maintains a higher potential for successful recognition. Second, voice detection is trained to differentiate human speech from environmental noise and the highest frequency range captured (6-8 kHz) may be unique to human speech played through a loudspeaker, and less likely to occur naturally in an environment. Lastly, there are certain consonants that are important for speech intelligibility that appear in the upper frequency range (~2-4 kHz) when recorded by a microphone~\cite{mic_univ}. The difference in frequency power within this range could also attribute to why some audio samples remain more intelligible (i.e., the samples with greater variance of power within that range).
\section{Summary and Discussion}
\label{sec:summary}

\noindent \textbf{Amazon vs. Google Observations:}
We observed some interesting trends between the two smart speaker devices that were used. Since Amazon and Google have their own speech processing services, it is reasonable to assume that different types of speaker samples will show different levels of success. If we consider the average wake word injection rates for both speaker types against the Amazon Echo Dot 2, we find that the TTS samples outperformed the live speaker recorded samples in all but one of the scenarios (Across-Door, 0.1 meters, 75 dB). Similarly, we find that the TTS samples outperformed the live speaker samples for command injection in all scenarios. This indicates that TTS samples are more effective for launching the \aname attack against Amazon devices. 

Another interesting observation that became apparent when comparing the injection success rates was that wake word injection was consistently more successful when attacking the Google Home mini device. By averaging the success rate of all speaker samples for each scenario, we find that there was more success at injecting the wake word to the Google device than the Amazon device across all scenarios that were tested. 
We also observed through our experiments that the Google device had significantly more instances of mis-recognized commands compared to the Amazon device. At lower SPL levels or larger distances where the Amazon device would simply disregard the audio that it heard, the Google device would make some attempt at recognition and provide some type of response, although often incorrect.

Lastly, our work demonstrates the feasibility of \aname when launched in scenarios without environmental noise. Naturally, this would be the most ideal setting to launch the attack and ensure no other audio in the environment interferes with the injection of the command. However, we believe that some environmental noise may be manageable and still allow for a successful attack. With the inbuilt noise cancellation capabilities of modern day VA devices, any environmental noise that is quieter than the injected command audio (after it passes through the barrier) will likely be filtered out by the device and the command will still be recognized.

\noindent \textbf{Sound Rating Values:}
We compare our observed results to the known Sound Transmission Class (STC) and Noise Reduction Coefficient (NRC) values for each of the barriers that we tested. We chose these rating values because they are both based on ASTM standards. 
If we consider the STC values for each of the barriers, we can see that our results are inline with the known values (33 for across-window, 30/34 for across-wall, and 20 for across-door). Now if we look at the NRC values for the different barrier materials, we find that both glass and gypsum board (i.e., drywall) have NRC values of 0.05 and wood has an NRC value of 0.10-0.15. All of these values are very low on the [0,1] scale indicating that none of the materials reflect much of the command audio back. 


\noindent \textbf{Drone-based Attack:}
Our drone experiments clearly demonstrate the feasibility of launching the \aname with drones. Specifically, an attacker could utilize a drone with a low operating loudness that does not impact the required SPL of the command audio to be injected. And by selecting the best performing command audio samples they may achieve up to 100\% command injection success. This method of launching the attack provides an attacker the benefits of remote command injection and the ability to target multiple (potential) victim devices in the same area without having to physically relocate or move their attack setup. Additionally, an attacker could utilize the Wi-Peep~\cite{hurley2022} exploit to initially locate the location of the target device before launching the attack.

\noindent \textbf{Improving Attack Robustness:}
From our experimental and analysis results we have deduced a few ways to increase the robustness of the \aname attack. First, while our results at 85 dB demonstrate the feasibility of the attack, using even louder command audio will increase the chances of attack success. An attacker can launch the \aname attack while the person is away from the home or they are in a situation where the loud audio will not cause detection. Higher volumes outside should not cause a problem, especially in scenarios with high rise buildings. An attacker could also plant a small wireless speaker onto a door or window that they could use to inject a command remotely. These devices can be very small and cheap~\cite{mpspeaker}, allowing the attacker to remain discrete.

Learning the type of voice assistant device that the user has before launching the attack would also help improve the chances of success. Our results demonstrate that different speaker types can have different levels of effectiveness for different devices. As a general observation, using TTS speaker samples would likely be the most effective for the \aname attack. Our analysis revealed that samples with stronger frequencies in the upper range are the most successful, so specifically choosing TTS samples that contain these qualities will improve attack success.


\noindent \textbf{Limitations:}
The results that we observed for the \aname attack are somewhat limited to the particular settings that we controlled in our experiments. Firstly, all of our experiments were conducted in quiet spaces where the only audio present was played from the loudspeaker device for the purposes of the attack. In a real-world scenario it is likely that there are other sources of noise in the environment which would affect the overall success of this attack. Second, since our attack uses plain, audible commands for the injection, the \aname attack is dependent on the user being away from the device and in another area. Otherwise they would easily recognize the command injection attempt. Lastly, as our results demonstrated, there is a distance requirement between the victim's device and the barrier (in the across-wall and across-window scenarios) for the attack to be successful at the SPL levels we tested. While louder command audio would surely increase the attack range, it also increases the chance of discovery. Therefore, the \aname attack is limited to scenarios where the victim's device is in close proximity to the barrier being targeted.

\noindent \textbf{Potential Defenses:}
The potential defenses against the \aname attack are largely based on hindering the physical phenomenon that would allow command audio to bypass physical barriers. 
One potential defense against our attack would be to use materials with higher STC and NRC values. To defend against the attack presented in this work, an STC of 50 or higher would be required. This can be achieved using concrete masonry walls, doubling the layers of drywall, or using specialized materials such as sound deadening paint or noise blocking curtains. Another potential defense is placing the smart speaker device at the furthest location from any accessible barriers. We demonstrate that distances of 4 meters become difficult for the attack even for an interior door. 
Another solution is to build a machine learning classifier that can differentiate between audio played through a barrier and audio played normally. As our analysis demonstrated, there are certain frequencies that are affected/blocked by the different barrier types. 
Blue et al.~\cite{blue_hello} achieve this effect by identifying sub-bass over-excitation which is a characteristic of audio played from loudspeaker devices and is not present in human speech. This would also be effective against \aname because as the command audio passes through the physical barrier, the bass/sub-bass components of the audio will become stronger. Another solution presented by Blue et al.~\cite{blue_2MA} could also be effective at identifying the \aname attack. In their 2MA work, the authors present a two microphone authentication solution that provides source localization by determining the direction of arrival. This approach, combined with a predetermined knowledge of the VA devices placement, could be used to identify when a command is coming from the other side of a barrier.
\section{Related Work}
\label{sec:relatedwork}

\noindent \textbf{Replay Attacks:}
Among all the spoofing attacks against VA systems, replay attacks are the most accessible to the adversary since it simply involves recording a victim's voice commands with a handy recording device and replaying the commands for the attack. Existing studies have shown that such attacks are effective against state-of-the-art speaker verification systems~\cite{Ergunay2015OnTV, Delgado2021ASVspoof2A}, under scenarios of replaying over the internet or within the physical space of the victim. Other than directly replaying the recorded speech, recent studies also reveal the potential ways of enhancing the stealthiness and effectiveness of the attack. VMask~\cite{Zhang2020VoiceprintMA} designed adversarial machine learning techniques to generate subtle perturbations to make any recorded speech pass speaker verification systems. To improve the stealthiness, Guo~\textit{et al.}~\cite{Guo2021ADM} exploited a loudspeaker array to make the sound emission focus on the microphone of the VA system, thereby hiding the attack sound from the user. To bypass existing defense schemes, Yoon~\textit{et al.}~\cite{Yoon2020ANR} leveraged a mouth simulator instead of a loudspeaker to replay the recorded speech. However, in terms of command injection attacks against smart speaker devices, replay attacks using commands in the victim's voice are not always necessary. Many of the current VA devices available (such as those used in our study) do not employ strict speaker verificatio. If the audio is understandable via speech recognition, the device will execute any command that is given which is a characteristic we exploit in \anametwo.
   
 

\noindent \textbf{Laser-based Injection:}
In addition to replay attacks, laser-based injection has also been utilized for signal and command injection targeting smart speakers. Recently, Light Commands~\cite{Sugawara2020LightCL} has brought up a new security issue, which is a new class of signal injection attacks targeted on microphones of the smart speakers by physically converting a light signal to sound signal. The attacker can inject arbitrary audio signals to a target microphone by aiming a specially designed amplitude-modulated light at the microphone’s aperture. By means of Light Commands~\cite{Sugawara2020LightCL}, the attacker can obtain control over some commodity smart speakers, such as Amazon’s Alexa, Apple’s Siri, and Google Assistant, at distances up to 110m, which provide a brand new perspective for attacking smart speakers. One drawback to this form of attack is that it requires a direct line-of sight between the attacker and victim's device. Therefore, simply closing the blinds or moving your device to a location out of view will thwart this attack. Our \aname attack does not require this line of sight and is much more accurate in practical settings.

\noindent \textbf{Ultrasonic/Hidden Audio Injection:}
In addition to conventional attacks through replaying human-sounding speech, researchers also show the potential of generating unintelligible or even inaudible attack sounds. Particularly, DolphinAttack~\cite{Yan2021TheFO} modulates the recorded voice commands onto the ultrasonic frequency range, which is inaudible to human hearing but can be demodulated by the microphone due to their non-linearity. Hidden voice attacks~\cite{Carlini2016HiddenVC, Abdullah2019PracticalHV} convert recorded speech into obfuscated voice commands, which are recognizable to the speech recognition models while remaining unintelligible to humans. Recent studies also demonstrate the possibilities of embedding such commands into background music~\cite{Yuan2018CommanderSongAS} or the audio channel of video streams~\cite{Zhou2019HiddenVC}. By combining hidden voice commands with live speech, the hybrid commands can even bypass the state-of-the-art defense schemes~\cite{Wu2021HVACEC}. While hidden voice commands introduce new approaches to evade detection, they are often very complicated to produce and are not feasible for real-world attack settings. Our attack does not obfuscate the command itself, but rather injects the clear-text command through a barrier. Hidden voice commands are obfuscated commands and hence they are often misrecognized or not effective. In a recent work by Abdullah et al.~\cite{Abdullah2021FaultIOA}, the authors survey current research works that present hidden voice command type attacks and demonstrate through experimentation that most of them will not be successful when launched against real-world systems. In this work we evaluate \aname against live implementations of VA devices and bypass real barriers with greater ease and feasibility than hidden voice commands.
 
\section{Conclusions and Future Work}
\label{sec:conclusion}

In this work, we present the \aname attack that issues audible voice commands to smart speakers across physical barriers. 
Our attack demonstrates the settings in which clean command injection can be successful and what barrier types are at risk.
This attack can be launched in person or remotely via drones or other controlled devices, and allow an attacker to gain full control over a victim's VA device when the device is placed near a barrier and the scenario allows for loud command audio to be played.
Compared to other command injection attacks, \aname exploits the lack of speaker verification present on modern smart speaker devices and bypasses physical barriers that would hinder other types of attacks. 
We evaluated the attack in multiple settings that test different command audio SPL levels and distances. Our experiments tested three different barrier-based attack scenarios using two live implementations of smart speaker devices and demonstrate that 100\% wake word and command injection accuracy can be achieved when selecting the highest performing speaker samples and under certain conditions. 

\newpage

\bibliographystyle{ACM-Reference-Format}
\bibliography{refs}

\appendix

\newpage
\onecolumn


\section{Appendix}

\subsection{Additional Images}
\label{app-imgs}

\begin{figure}[h]
\centering
\begin{minipage}{0.35\textwidth}
	\raggedright
	\captionsetup{font=footnotesize,labelfont=bf}
	\centering
	\subfloat[No Insulation]{
		\includegraphics[scale=0.0335]{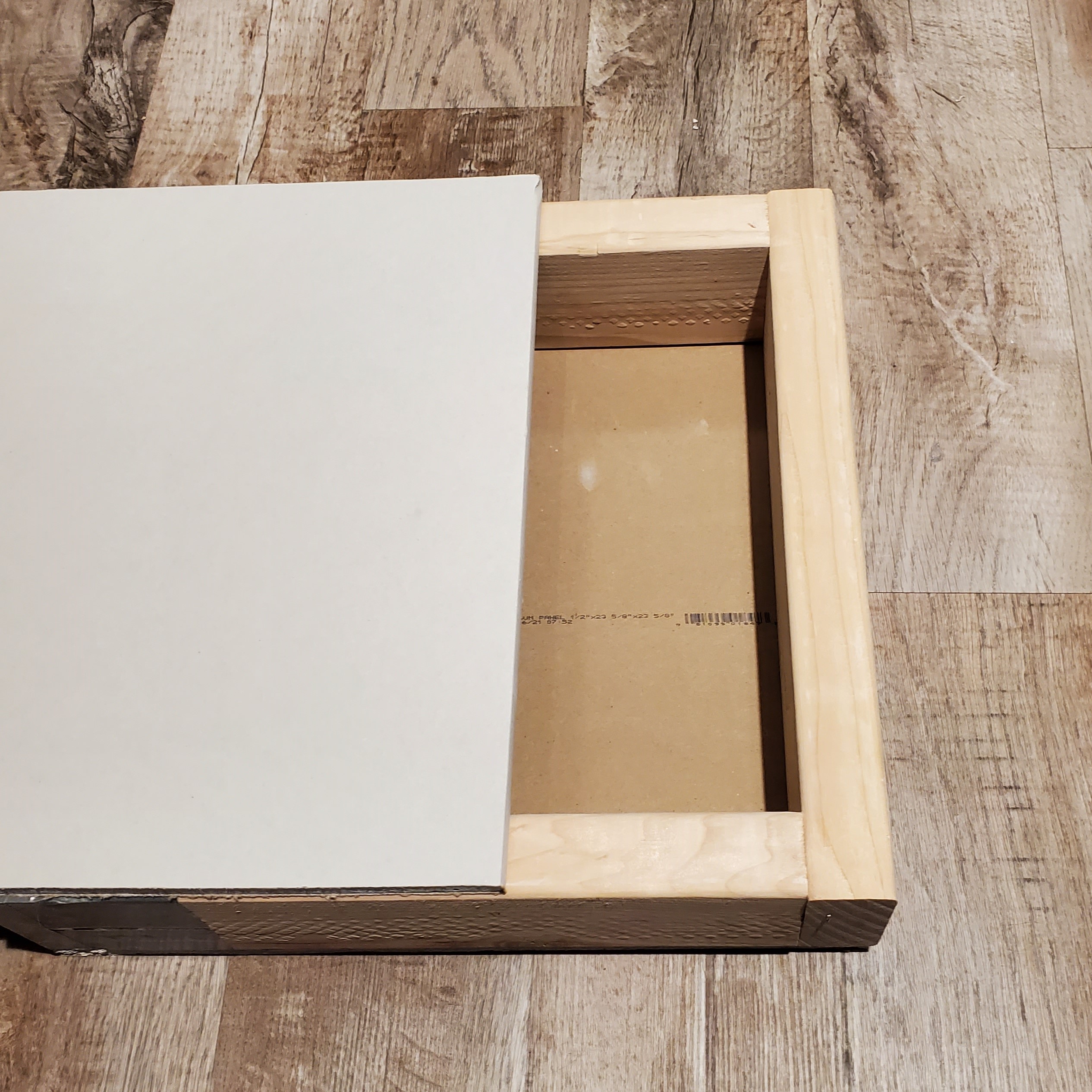}
	}
	\subfloat[Insulated]{
		\includegraphics[scale=0.032]{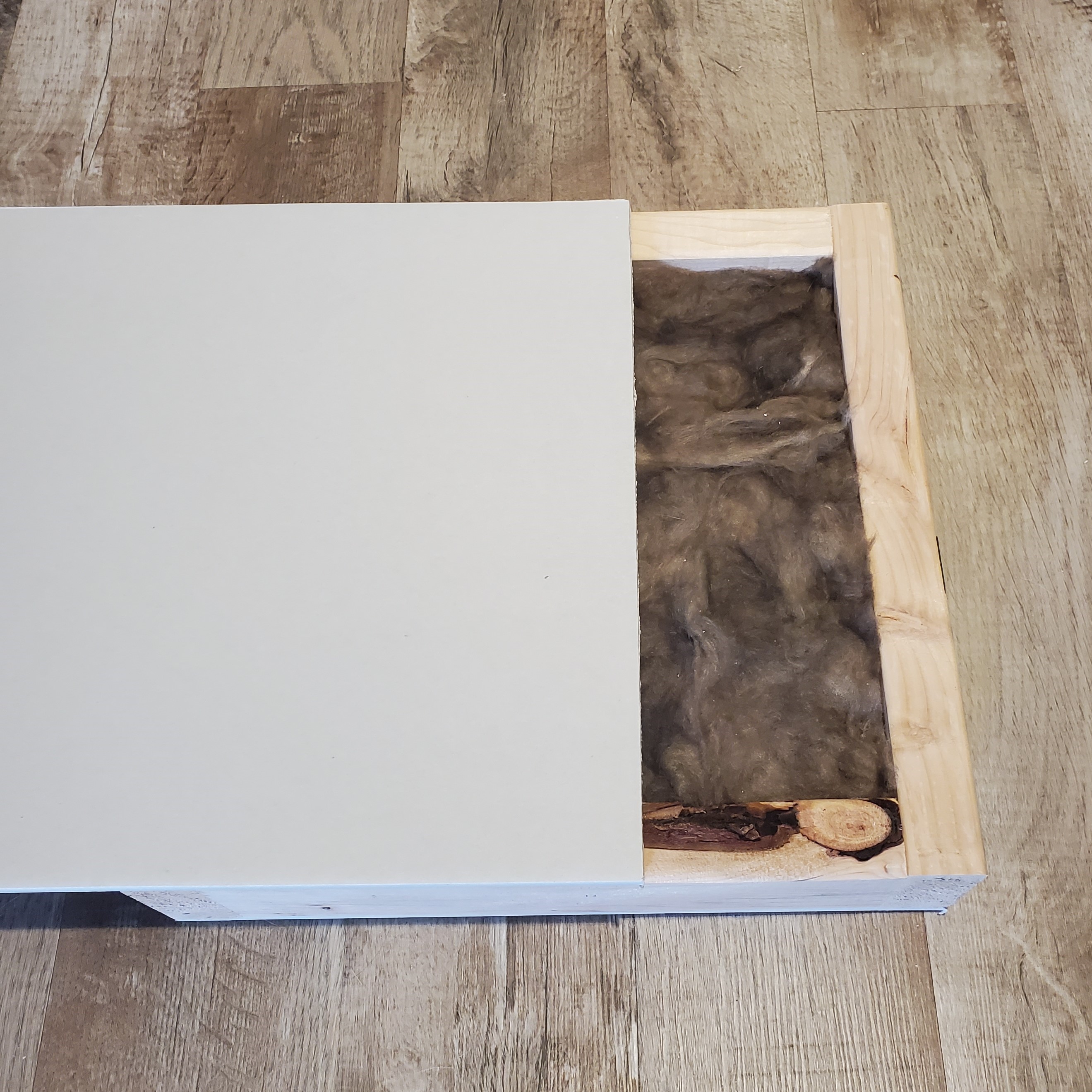}
	}
	\caption{Inserts constructed for the Wall-Barrier.}
	\label{fig:inserts}
\end{minipage}%
\begin{minipage}{0.65\textwidth}
	\raggedleft
	\captionsetup{font=footnotesize,labelfont=bf}
	\subfloat[Soundproof Box]{
		\includegraphics[scale=0.0305]{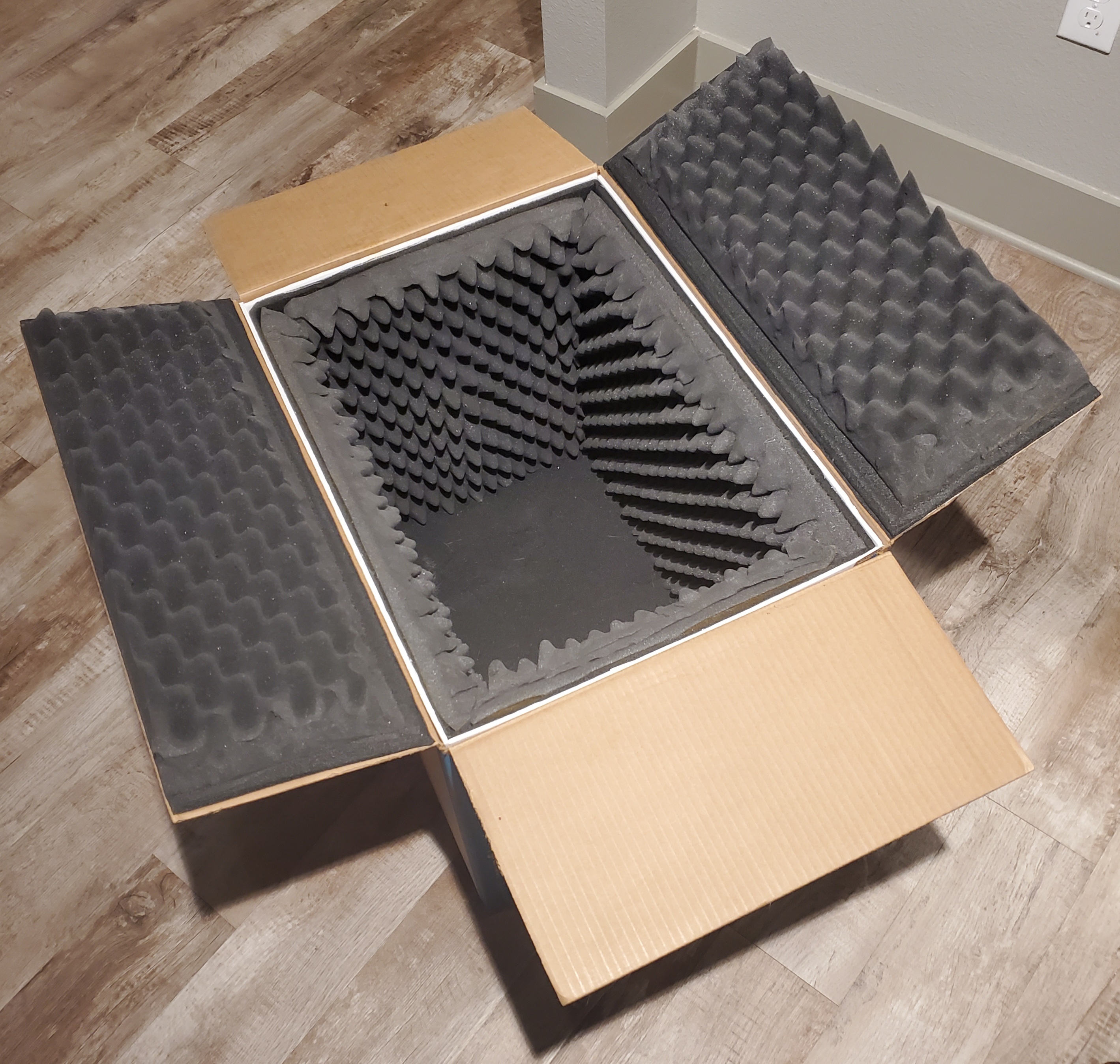}
		\label{subfig:box}
	}
	\hspace{0mm}
	\subfloat[Box Layers]{
		\includegraphics[scale=0.165]{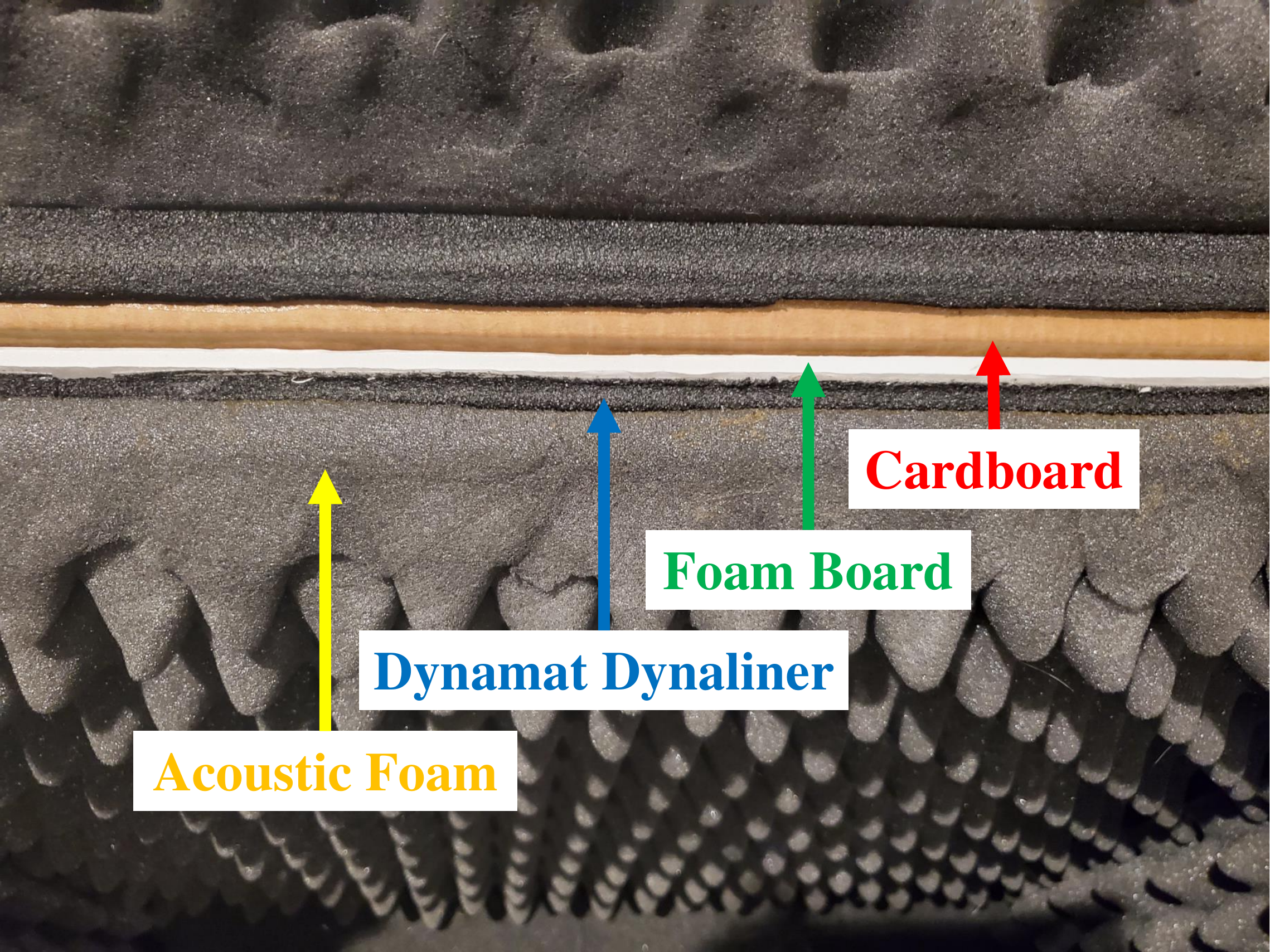}
		\label{subfig:layers}
	}
	\subfloat[Experiment Aerial View]{
		\includegraphics[scale=0.15]{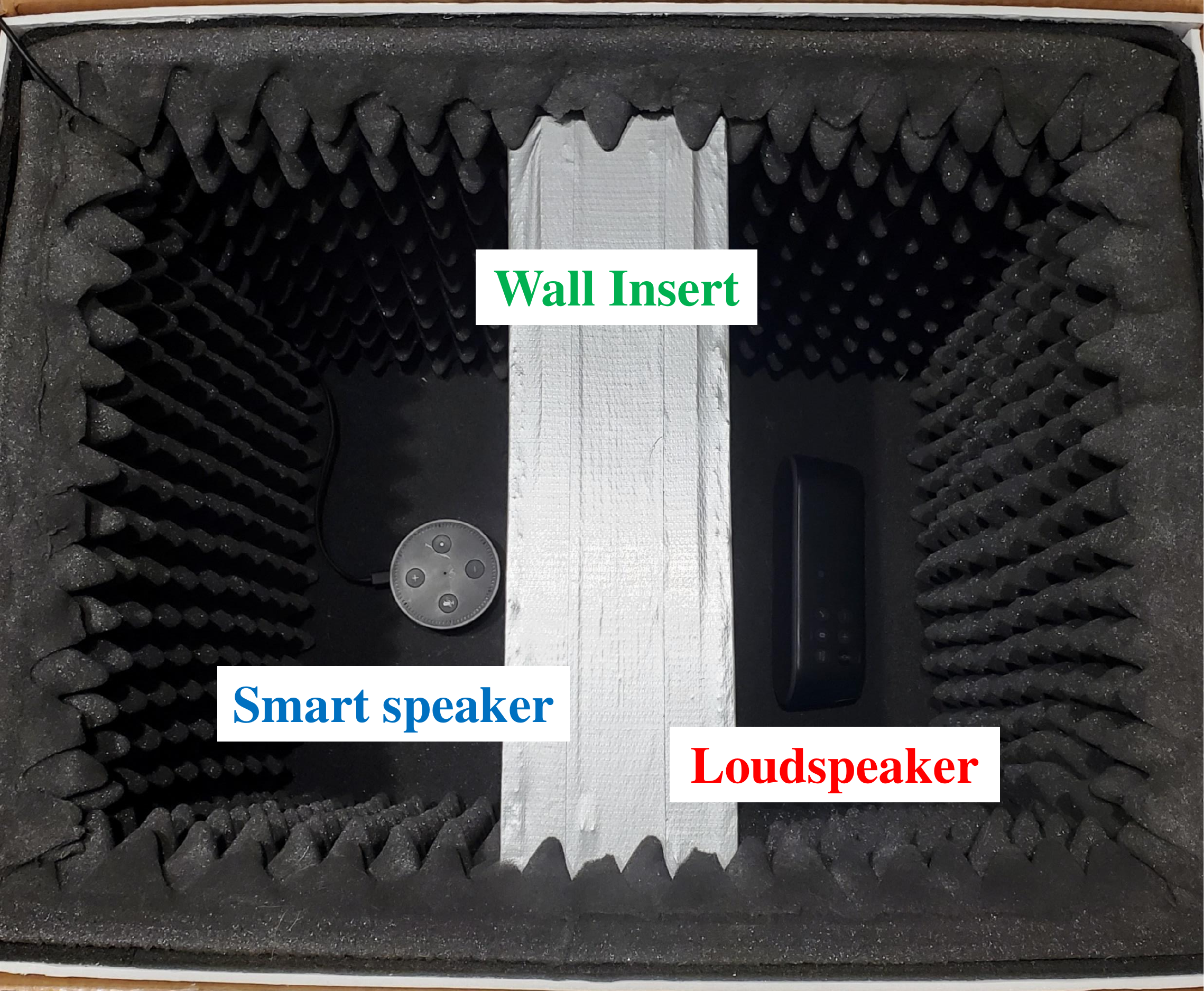}
		\label{subfig:exp_aerial}
	}
	\caption{Images of soundproof box construction and experimental setup.}
	\label{fig:sp-box}
\end{minipage}
\end{figure}

\subsection{Additional Tables}
\label{app-tables}

\begin{table*}[h]
	\captionsetup{font=footnotesize,labelfont=bf}
	\centering
	\caption{Wake Word injection success rates, for attacking the Amazon Echo Dot 2, for each Barrier scenario. *Table is condensed to include only rows that showed some injection success.}
	\includegraphics[scale=0.66]{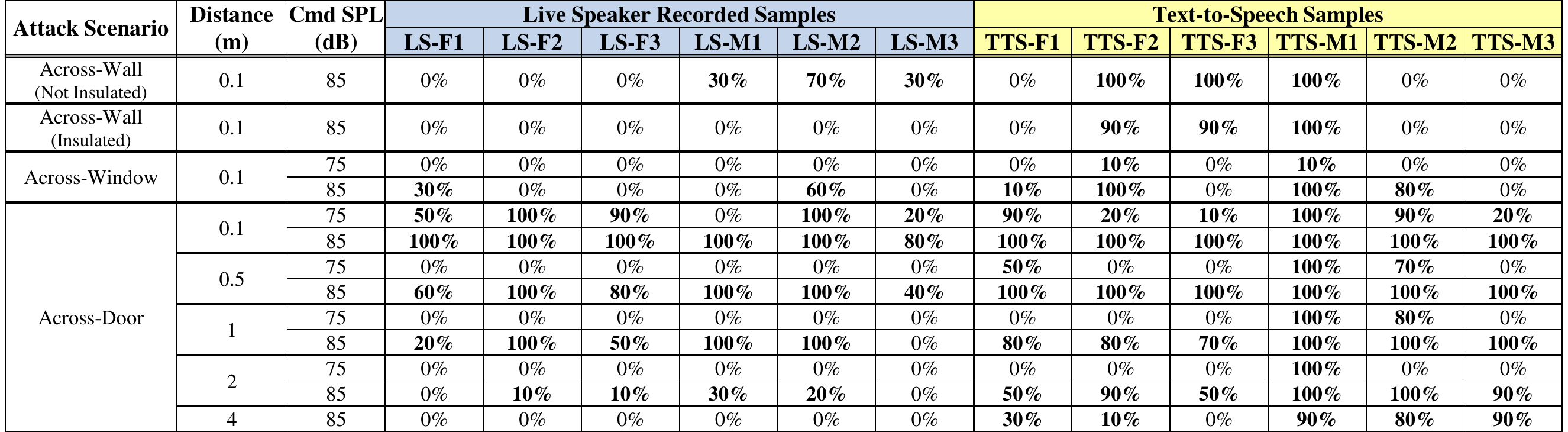}
	\label{tab:alexa-attack-results_ww}
\end{table*}
\begin{table*}[h]
	\captionsetup{font=footnotesize,labelfont=bf}
	\centering
	\caption{Wake Word injection success rates, for attacking the Google Home mini, for each Barrier scenario. *Table is condensed to include only rows that showed some injection success.}
	\includegraphics[scale=0.66]{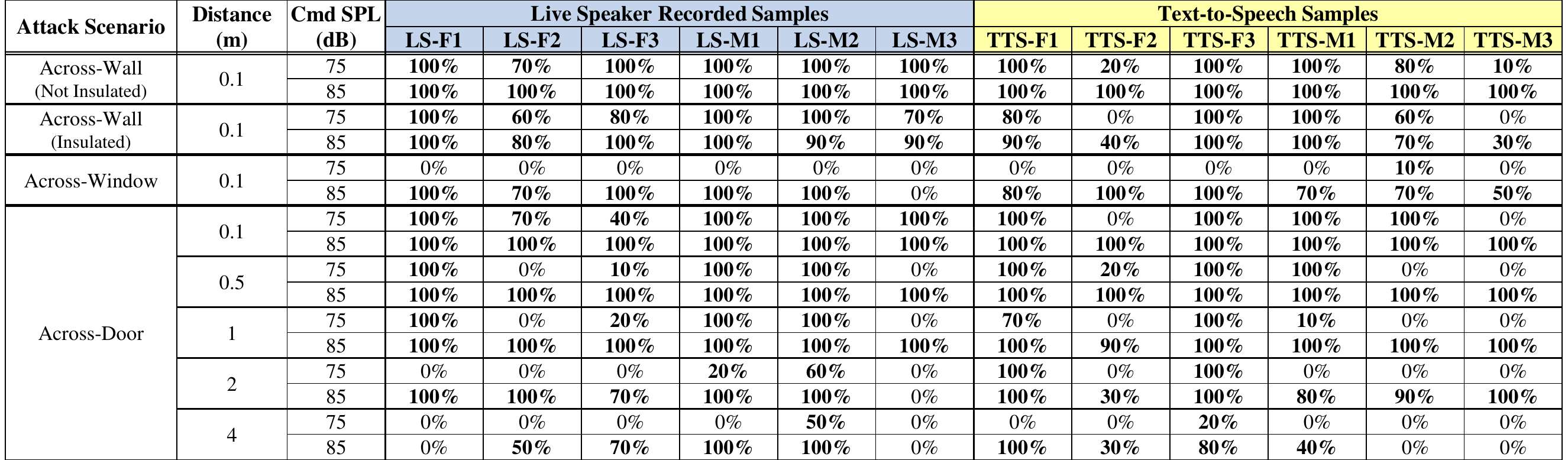}
	\label{tab:google-attack-results_ww}
\end{table*}

\end{document}